# Phase-matched nonlinear wave-mixing processes in XUV region with multicolor lasers

**Khoa Anh Tran,\* Khuong Ba Dinh, Peter Hannaford and Lap Van Dao**

*Centre for Quantum and Optical Science, Swinburne University of Technology, Melbourne 3122, Australia*

*\*Corresponding author: anhkhoatran@swin.edu.au*

**Abstract**

We report here experimental results of perturbative nonlinear optical wave-mixing processes in the extreme ultraviolet region by using two-color and three-color laser fields. Besides the usual odd-harmonic spectrum of high harmonic generation, new spectral components are observed when multiple incommensurate lasers (one driving plus one or two control field) interact with neutral krypton gas. To demonstrate the wave-mixing process underlying such an observation, we firstly couple the driving field with either the signal or the idler field of an optical parametric amplifier in the gaseous ensemble to generate certain mixing frequencies. The two control fields are then simultaneously combined with the driving field to produce broad and distinguishable mixing peaks that clearly reveal the contribution of each control laser. Finally, the variation of the intensity of the mixing waves with the intensity of each control field, the gas density, and the relative focus position is examined for signatures of phase-matched generation of the mixing fields in this spectral region.

## 1. Introduction

Recent breakthroughs in ultrafast optical physics [1] allow us to study laser-matter interactions in the strong-field regime, and the quasi-classical model [2] and quantum mechanical calculations [3] for the high harmonic generation (HHG) process make recollisional physics a promising route for the synthesis of isolated attosecond pulses. The phase-matched HHG process produces electromagnetic waves with high spatial [4] and temporal [5] coherence. The ultrashort pulse duration and ultrahigh photon frequency [6, 7] of these light sources thus play a pivotal role in attosecond transient absorption spectroscopy [8–11], in studies of configuration interactions [12], and in gaining structural information of the emitting media [13, 14].

In HHG, a bound electron is submitted to a non-perturbative process which consequently emits photon energy in the soft X-ray region. The macroscopic HHG spectrum possesses a comb structure of equidistant and quasi-constant-strength narrow line shapes. Absorption spectroscopy techniques requiring precise wavelength coherent light fields, however, limit the versatility of HHG sources. Thanks to the high nonlinearity underlying the HHG process, new mixing frequencies were successfully extended to the extreme ultraviolet (XUV) and soft X-ray regimes with a two-color gating method [15–22]. Desired radiations that are inaccessible with pure HHG have thus become accessible with phase-matched cascaded wave-mixing processes in the XUV region [21, 22]. Additionally, the phase of the generated harmonics can also be controlled by a delayed laser field [23]; hence the cutoff energy, the intensity, and the spectrum of HHG are enhanced.

In contrast to HHG, the intensity of low-order harmonics with perturbative nonlinear optical processes varies linearly with the intensity of the driving field; the higher the harmonic, the lower the output spectral power [24]. Recently, perturbative nonlinear optics under extreme conditions has been developed by extending conventional nonlinear optics to the XUV and soft X-ray region [21, 22]. In the multiphoton picture, the energies of the mixing waves gated by the two-color HHG (strong field $\omega_1$ interfering with a weak field $\omega_2$), where $n_1$ and $n_2$ are the numbers of photons of each field $\omega_1$ and $\omega_2$, and $n_1 + n_2$ must be an odd number [18]. With an arbitrary frequency ratio $\omega_1/\omega_2$ (commensurate two-color synthesized field is a special case [15–18]), the energies of these mixing frequencies can also be read as $\omega_{mix} = q\omega_1 \pm n(\omega_1 - \omega_2)$, where $q$ is the harmonic order [21, 22]. Consequently, the desired energies of the mixing peaks relative to the odd harmonics can be obtained by using an appropriate frequency for the weak control field $\omega_2$. From a general perspective, the quasi-continuous mixing frequencies $\omega_{mix}$ that further narrow the spectral gap between two odd-harmonics can be produced with the use of multiple control lasers. Additionally, the macroscopic phase mismatch arising from the wave-mixing processes is one of the traditional concerns for the production of intense coherent XUV pulses.

In this paper, we study phase-matched perturbative nonlinear optical processes in the XUV region using collinear two-color and three-color laser fields. Besides the sharp HHG spectrum driven by an 800-nm laser (driving field), non-integer-order wave-mixing spectra are produced when the driving field and one control field (either the signal field at 1400 nm or the idler field at 1860 nm) of the optical parametric amplifier (OPA) are focused into a krypton medium. Moreover, the simultaneous presence of three laser fields of 800 nm, 1400 nm and 1860 nm generates resolvable four-wave mixing (FWM) frequencies that signify the contribution of each control field. The intensities of the mixing frequencies with two-color lasers are also investigated as a function of the intensity of each control field, the krypton gas pressure, and the focus position for clear evidence of phase-matched FWM processes in this spectral region. Thus, our experimental results provide a potential method using multicolor phase-locked control lasers to induce an intense XUV quasi-continuum spectrum for the synthesis of isolated attosecond pulses and for ultrafast absorption spectroscopy.

## 2. Nonlinear optics in the extreme ultraviolet
High harmonic generation is stimulated when high-intensity laser pulses interact with a neutral medium; the gradient force arising from the strong laser's electric field excites atomic charges from the ground state to the continuum. The induced atomic dipole oscillations are consequently source terms for the generation of new harmonic frequencies as the accelerated free-electron wave packets return to the nucleus. The generated harmonic waves and the fundamental field, however, experience a number of sources of dispersion due to macroscopic propagation through the interaction medium. In this study, the driving laser beam is weakly focused with a lens ($f = 25$ cm) so that the radius of the focal volume is approximately 75 $\mu$m over which the Rayleigh length is ~10 mm. The HHG field is produced with 25 Torr of krypton gas at an effective intensity of ~1.2 x $10^{14}$ W/cm$^2$. The nonlinear refractive index $n_2$ of krypton at 1 atm is ~2.8 x $10^{23}$ m$^2$/W [25, 26]; $n_2I$ is thus ~7% of the linear refractive index at an intensity of ~2.0 x $10^{14}$ W/cm$^2$ [27]. Thus, the effect of the nonlinear refractive index is negligible. The focus position of the lens (the center of the Rayleigh length) is shifted from outside to inside of the gas cell so that the interaction length for coherent accumulation of the output harmonic source can be optimized. The effective interaction length, which is estimated from the displacement of the focus position over which the harmonic intensity scales quadratically and therefore the phase-matching condition of the HHG is satisfied [28], is ~4 mm. This is shorter than the Rayleigh length so that the true phase-matching condition for the HHG field can be established when the dispersions of the neutral atoms and the free-electron plasma cancel each other regardless of the dispersions originating from the interaction configuration (Gouy phase shift) and the harmonic dipole phase [21, 29]

$$\Delta k_q = k_q - qk_l \approx -\frac{2\pi q}{\lambda_l}P\delta(\lambda_l)(1-\eta) + P\eta N_{atm}r_e\lambda_l\left(q-\frac{1}{q}\right) \approx 0. \qquad (1)$$

In Eq. (1), the first and the second terms correspond to the dispersions of the neutral atoms and the free-electron plasma; $\lambda_l$, $P$, $\delta(\lambda_l)$, $\eta$, $N_{atm}$, and $r_e$ denote the wavelength of the driving laser, the gas pressure, the neutral dispersion of the fundamental field, the ionization fraction, the gas density at one atmosphere, and the classical electron radius, respectively. The frequency of the $q$th harmonic is greater than the ionization potential; the neutral dispersion of the harmonic field through the bound-free transitions is more than two orders of magnitude lower than that of the fundamental field ($\delta(\lambda_l) >> \delta(\lambda_q)$) [30]; thus the contribution of the $\delta(\lambda_q)$ is negligible. It is worth noting that we use a low krypton gas pressure (25 Torr) as the interaction medium and the ionization rate is low (<4%) [31]. The dispersion arising from generated free-electron plasma is effectively counterbalanced by the neutral dispersion. Thus, the plasma defocusing effect is also negligible in this study. Under such conditions, the intensity of the $q$th-harmonic $I_q$ scales quadratically with the gas pressure and the effective interaction length [32, 33]. The use of a second weak control field in combination with a strong first field (which drives the HHG) reveals the possibility of developing a perturbative theory for the XUV region. In the weak-field limit of the control field, the generated HHG photons are likely to interact with the two fundamental lasers. Thus, the integrated intensity of the coherent nonlinear optical wave-mixing field in the XUV region gated by three electromagnetic fields over an interaction length $L$ can be written as [21, 22]

$$I_{mix} \sim I_q I^n_1 I^n_2 |\chi^{(2n+1)}|^2 N^2 L^2 \text{sinc}^2\left(\frac{\Delta k_{mix}L}{2}\right), \qquad (2)$$

where $I_1 \equiv I_{800}$ and $I_2 \equiv I_{1400}$ or $I_{1860}$ are the intensities of the driving field and the control field (the signal or the idler field), respectively. $n$ is equal to 1 or 2 corresponding to the number(s) of involved photon(s) of each fundamental field. The phase mismatch for the generation of these mixing frequencies is $\Delta k_{mix} \cong k_{mix} - k_q \pm n(k_1 - k_2)$, where $k_1 \equiv k_{800}$, $k_2 \equiv k_{1400}$ or $k_{1860}$, $k_q$ and $k_{mix}$ are the wave-vectors of the driving field, the control field, the $q$th-order harmonic and the mixing field, respectively. Within the framework of our study (constant $I_1$ and therefore $I_q$ is also unchanged), these perturbative processes are expected to follow the power-scaling of the weak control field $I^n_2$. Also, the dependence of the intensity of the mixing fields on the square of the gas density and on the square of the interaction length provides a straightforward way to investigate whether the accumulation of the mixing frequencies is induced with the macroscopic phase-matching condition.

## 3. Experimental setup

A 1-kHz chirped-pulse amplifier system provides 6.0 mJ, 800 nm, 30 fs laser pulses. The beam is divided into two independent optical paths with a 30:70 beam splitter. The first beam of energy ~4.0 mJ is adjusted to ~0.43 mJ (driving field, $\omega_1$) to produce HHG ($\omega_q \equiv q\omega_1$) in a 20-cm long cell filled with 25 Torr of krypton gas. The second beam of energy ~2 mJ pumps a commercial two-stage OPA (Quantronix, Palitra) to create 0.6 mJ energy of two NIR fields (control fields, $\omega_2$). The OPA is optimized for parametric amplification of the signal (central wavelength 1400 nm, 60 fs) and the idler (centred at 1860 nm, 60 fs) fields. The intensity ratio of the idler to the signal wave is about 1:10. The HHG is generated by ~1.2 x $10^{14}$ W/cm$^2$ 800-nm pulses and the control fields are weak, i.e., $I_{max}(1400) < 5$ x $10^{13}$ W/cm$^2$ and $I_{max}(1860) < 5$ x $10^{12}$ W/cm$^2$, so that they do not create HHG radiation by themselves. A half-wave plate is used to align the polarization of the driving field to that of the control fields. The signal and the idler waves are polarized perpendicularly to each other; therefore a linear polarizer is used to optionally transmit each control field (or both of them) in order to meet a certain condition. The intensity of the control field is varied with neutral density filters. The co-propagating driving and control beams are combined with a dichroic mirror before being focused into the gas cell by a focusing lens ($f$ = 25 cm). The phase-matching condition of the HHG is maintained during the experiment. A home-built telescope formed by a pair of lenses is installed in the control beam path to manipulate the spatial overlap of the driving and the control fields, and the synchronized delay time between these two fields is controlled by a DC motor with 0.1 fs temporal resolution. The focal beam spot is ~150 $\mu$m in diameter. The neutral krypton gas in the cell (glass entrance, aluminium rear surface) is isolated from the vacuum outside of the cell; a small pinhole (~150 $\mu$m) on the exit plane of the gas cell, from which the HHG generated in the cell propagates toward the CCD camera, is directly drilled by the focused 4-mJ 800-nm beam. The residual fundamental beams after emanating from the gas cloud are then blocked by a 300-nm-thick aluminium foil. The emitted XUV radiation is recorded with a flat-field 300 lines/mm diffraction grating and an X-ray CCD camera. Negative delay time implies that the control pulses precede the 800-nm driving pulses. More details of the experimental setup can be found elsewhere [22].

## 4. Results and Discussion
### 4.1. Perturbative nonlinear optics in the extreme ultraviolet region

The HHG spectrum at -300 fs delay spanning from H13 (13th harmonic) to H27 (27th harmonic) at a pressure of 25 Torr of krypton gas is shown in the inset (a) of Fig. 1. The HHG is induced with the phase-matching conditions ($\Delta k_q \approx 0$) which will be discussed in the following section. An intense and narrow-bandwidth harmonic comb is observed when the intensity of the 800-nm driving laser is ~1.2 x $10^{14}$ W/cm$^2$ and the focus position of the lens is ~3 mm inside the gas cell. The control field does not affect the HHG process at such a large negative time delay. Perturbative nonlinear optical processes in the XUV region are then gated with two-color lasers in which the OPA signal field at 1400 nm or the OPA idler field at 1860 nm is the control field. As the signal and idler waves are polarized perpendicularly to each other, the linear polarizer and the half-wave plate are manipulated so that the relative polarization planes between the 800-nm and 1400-nm fields, and between the 800-nm and 1860-nm fields, are parallel. Besides the odd-order HHG spectrum, non-integer-order harmonics are recorded when either the signal field or the idler field is temporally and spatially synchronized with the 800-nm field in the krypton gas, Fig. 1.

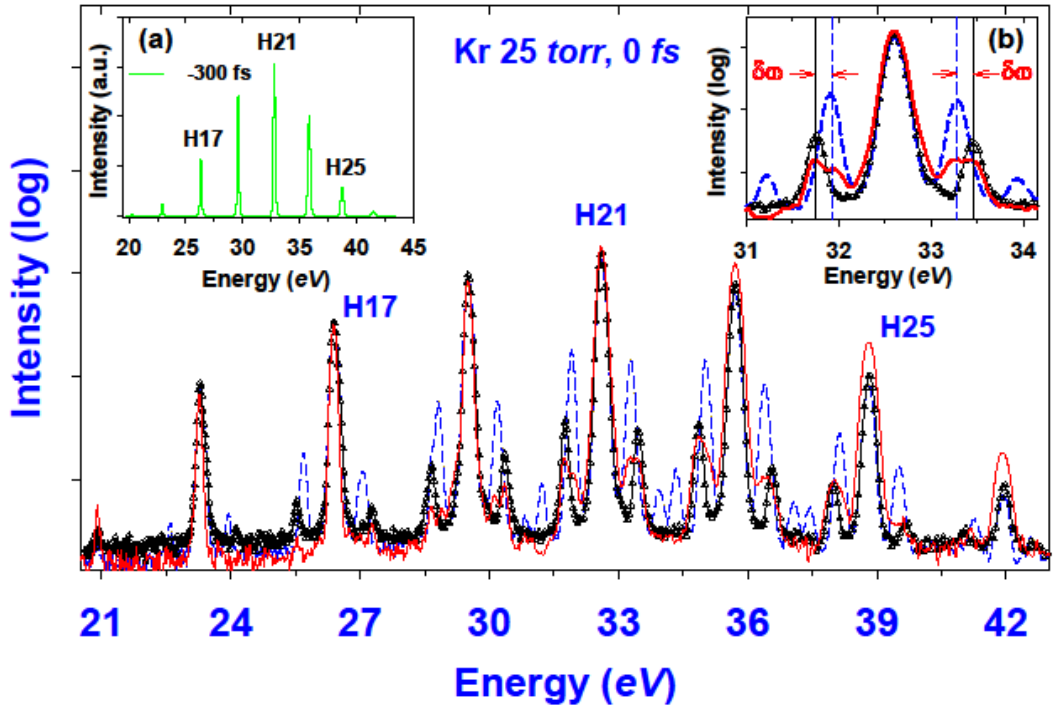

**Figure 1**. Wave-mixing spectra of the 25-Torr krypton gas at 0 fs time delay gated by two-color fields (800 nm and 1400 nm [blue dashed line], 800 nm and 1860 nm [black triangles]), and three-color fields 800 nm, 1400 nm, and 1860 nm (red solid line). Inset (a) shows the sharp HHG spectrum driven by the strong 800-nm field taken at -300 fs. Inset (b) shows the 21st harmonic and the corresponding mixing frequencies. Photon energy of the OPA signal field $\omega_{1400} \cong 0.89$ eV, OPA idler field $\omega_{1860} \cong 0.67$ eV; $\delta\omega \approx 0.24$ eV $\approx \omega_{1400} - \omega_{1860}$.

The experimental conditions for the generation of these new frequencies are maintained so that the sharpness and bandwidth of the HHG spectrum at 0 fs are almost the same as those at -300 fs. Such a condition is obtained by applying weak control fields, i.e., $I_{1400} < 5 \times 10^{13}$ W/cm$^2$ and $I_{1860} < 5 \times 10^{12}$ W/cm$^2$. With parallel polarization of the driving and control beams, the strongest mixing peaks are manifested at 0 fs time-delay between the 800-nm and 1400-nm fields (blue dashed line) and between the 800-nm and 1860-nm fields (black triangles). Due to the higher applied intensity, the two-color fields of 800 nm ($\omega_1 \equiv \omega_{800}$) and 1400 nm ($\omega_{1400}$) produce more mixing frequencies than do the 800 nm and 1860 nm ($\omega_{1860}$). Additionally, the mixing fields induced by the 800-nm plus 1400-nm fields are more intense than those induced by the 800-nm plus 1860-nm fields. The energy and momentum conservation in the perturbative approach leads to the generation of mixing fields of frequency $\omega_{mix} = \omega_3 \pm n(\omega_1 - \omega_2)$ and wave-vector $k_{mix} = k_3 \pm n(k_1 - k_2)$ where $\omega_3$ and $k_3$, $\omega_1$ and $k_1$, $\omega_2$ and $k_2$ are the carrier frequencies and the wave-vectors of the XUV, the 800 nm, and the 1400/1860 nm fields, respectively. The HHG field ($\omega_q \equiv \omega_3$ and $k_q \equiv k_3$), which is the XUV field in the wave-mixing processes, is generated under the phase-matching conditions. On one side of each odd harmonic, the energy differences between the odd harmonic and the nearest peak (first mixing-order) and between the first mixing-order and a further peak (second mixing-order) induced by the driving and the OPA signal fields are about 0.67 eV and 2x0.67 eV, respectively (blue dashed line, Fig. 1). These two energy values are approximately $\Delta\omega$, and $2\Delta\omega$, where $\Delta\omega = \omega_{800} - \omega_{1400}$. In the same way, the energies of the mixing waves gated by the driving and the OPA idler fields can be deduced with this rule. In fact, the energy separation between one odd harmonic and the first-order mixing frequency is ~0.89 eV (black triangles, Fig. 1). Such an energy gap is roughly equal to $\Delta\omega' = \omega_{800} - \omega_{1860}$.

In the inset (b) of Fig. 1, the harmonic H21 and mixing fields are shown for a better visualization of the energy calibration. The energy gap between the two vertical lines is $\delta\omega \approx \omega_{1400} - \omega_{1860} \approx 0.24$ eV. Thus, the general frequencies of the mixing fields induced by the mixture of the 800-nm and the 1400-nm fields (800-nm and 1860-nm) can be summarized as $\omega_{mix} = \omega_q \pm n\Delta\omega$ ($\omega_q \pm \Delta\omega'$), where $n$ is equal to 1 or 2. This expresses the energy conservation the of sum-frequency mixing and difference-frequency mixing processes in the XUV region. The phase variation of the fundamental field in the presence of free electrons is transferred to the harmonic field as a result of the nonlinear process of harmonic generation. The influence of the plasma and neutral dispersions on the propagation of the XUV fields is insignificant [29]; therefore the harmonic phase is almost constant over the coherence length. Because the gas pressure is low (25 Torr), the dispersion of $k_1$ is similar to that of $k_2$ and so the mismatch $\Delta k_{mix}$ is approximately zero for the mixing fields. In other words, the phase-matching condition is also fulfilled for the frequency-mixing processes in this spectral range. As will be discussed in the following section, the phase mismatch product $\Delta k_{mix} L \ll 1$ is maintained over a maximum length $L_{max}$ of about 2 mm where the mixing fields accumulate coherently.

In another measurement, we study the optical wave-mixing in the XUV region with a collinear scheme of three fundamental laser fields. The experimental conditions for the generation of HHG with a 800-nm field is the same as that discussed above. The two control fields at 1400 nm and 1860 nm are now allowed to be partially transmitted through the linear polarizer when the polarizer's transmission axis is not parallel with the polarization directions of the control fields. Owing to a much weaker intensity of the idler field relative to that of the signal field, the idler field is favored by aligning the linear polarizer so that the polarizer's transmission axis and the polarization plane of the idler field forms a small angle (~15º). With such an alignment, the presence of three fields of 800 nm, 1400 nm and 1860 nm consequently introduces a broader mixing spectrum (red solid line, Fig. 1). This spectrum clearly signifies the contribution of each control field, i.e., two distinguishable FWM frequencies on one side of each odd harmonic. The energies of these two peaks are exactly equal to that of the FWM fields produced by the two-color laser fields (see inset (b) of Fig. 1). However, the spectral power of the resultant mixing fields is much weaker than that of the FWM fields generated with the parallel two-color lasers.

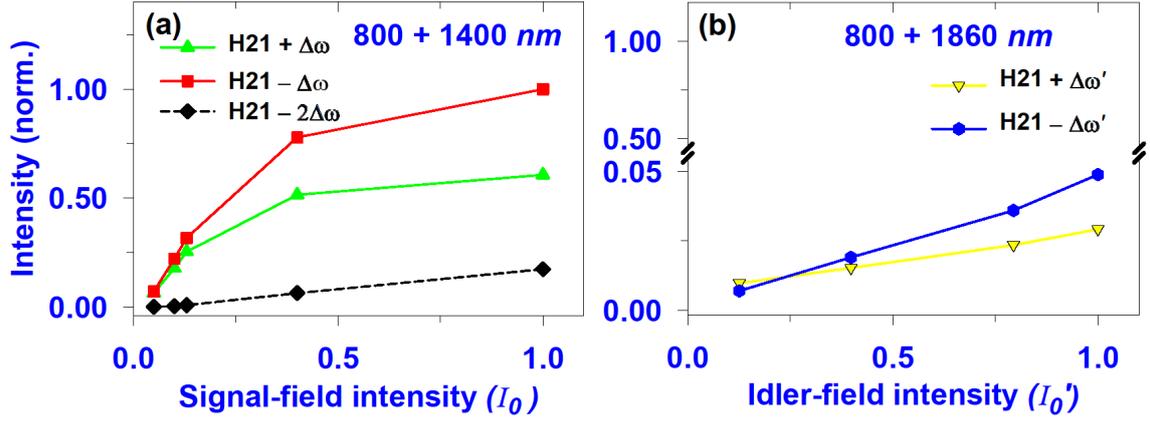

**Figure 2**. (a) Dependence of the intensity of H21 + $\Delta\omega$ (green triangles), H21 - $\Delta\omega$ (red squares), and H21 - 2 $\Delta\omega$ (black diamonds) on the relative intensity of the 1400-nm signal field. (b) Dependence of H21 + $\Delta\omega'$ (yellow triangles), and H21 – $\Delta\omega'$ (blue hexes) on the intensity of the weak 1860-nm idler wave. $I_0 < 5 \times 10^{13}$ W/cm$^2$, $I'_0 < 5 \times 10^{12}$ W/cm$^2$, $\Delta\omega = \omega_{800} - \omega_{1400}$, and $\Delta\omega' = \omega_{800} - \omega_{1860}$. The vertical normalized scales are shown relative to (a).

The optical wave-mixing processes are induced by different fields, and therefore the intensities of the generated waves are expected to agree with Eq. (2). Within the control-field-intensity dependence framework, $I_1$ (and therefore $I_q$) is maintained whilst $I_2$, i.e., $I_{1400}$ or $I_{1860}$, is varied with neutral density filters. We only discuss experimental data taken with a parallel polarization state between the driving and control fields at 0 fs time delay. The maximum intensity of the signal (idler) field $I_0$ ($I'_0$) is below 5 x 10$^{13}$ W/cm$^2$ (5 x 10$^{12}$ W/cm$^2$). For the first mixing-order [n = 1, Eq. (2)] when $I_{1400} < 0.2I_0$, the intensity of the H21 + $\Delta\omega$ (green triangles) and the H21 - $\Delta\omega$ (red squares) increases as a linear function of $I_{1400}$, Fig. 2(a). When $I_{1400} > 0.2I_0$, the increment of the H21 ± $\Delta\omega$ fields behaves in a nonlinear way and finally saturates. Such a behavior could be due to a gradual decrease of the observed H21 intensity (and therefore a decrease in $I_q$ in Eq. (2)). The photon yield of H21 - 2$\Delta\omega$ (black diamonds), however, increases linearly over the full range of $I_{1400}$. Additionally, the intensity of the FWM fields induced by the driving and the idler field also seemingly obeys Eq. (2), i.e., a monotonic rise of the power of H21 + $\Delta\omega'$ (yellow triangles) and H21 - $\Delta\omega'$ (blue hexes) up to the maximum intensity $I'_0$ of the idler field, Fig. 2(b).

We are unable to fully implement a power-scaling measurement of the second mixing-order waves in this study since many complicated nonlinear processes (which destroy the sharp spectra of the HHG and the mixing frequencies) are involved when $I_{1400} \gg I_0$. However, the temporal signal decay of the second mixing-order could provide indirect evidence of the nonlinear processes involving the $\chi^{(5)}$ susceptibility under our experimental conditions [21]. Spectral powers of other harmonics and mixing frequencies behave in a similar manner to those discussed above. In summary, the conservations of energy and momentum, the appearance of the same four-wave mixing frequencies in two cases of the two-color, and three-color lasers, the linear power-scaling of the four-wave mixing frequencies with respect to the control fields are clear evidence of perturbative nonlinear optical wave-mixing processes in the XUV region.

## 4.2. Phase-matched four-wave mixing in the extreme ultraviolet

The krypton gas pressure or the focus position of the lens is varied so that effects of the phase mismatch on the mixing waves with a parallel polarization state between the two-color fields can be examined. The experimental condition for HHG with the 800-nm field at -300 fs is the same as that described in Section 4.1. In the study of the pressure dependence, all experimental parameters are unchanged except the gas pressure, which is varied from 5 Torr up to 150 Torr. As a result, the intensities of the FWM frequencies, i.e., H21 + $\Delta\omega$ (green diamonds), and H21 - $\Delta\omega$ (blue squares) at 0 fs, gated by the two-color fields of 800 nm and 1400 nm closely follow that of the harmonic H21 taken at -300 fs (black circles) and 0 fs (red triangles), Fig. 3(a). Similarly, the intensity profiles of the H21 + $\Delta\omega'$ (green crosses) and H21 - $\Delta\omega'$ (blue pluses) produced by the mixture of 800 nm and 1860 nm fields also follow that of the main harmonic H21, Fig. 3(b). The increment of the intensity of the H21 (and therefore all mixing fields) exhibits the $p^2_{Kr}$-dependence when $p_{Kr} \leq 25$ Torr (region between the two red arrows). This is a typical aspect of phase-matched generation of HHG radiation [22, 31, 32]. When $p_{Kr} > 25$ Torr, all observed frequencies decrease in intensity due to a large phase-mismatch. At pressures above 40 Torr the dominant re-absorption effect of the krypton gas results in an exponential decay of all fields.

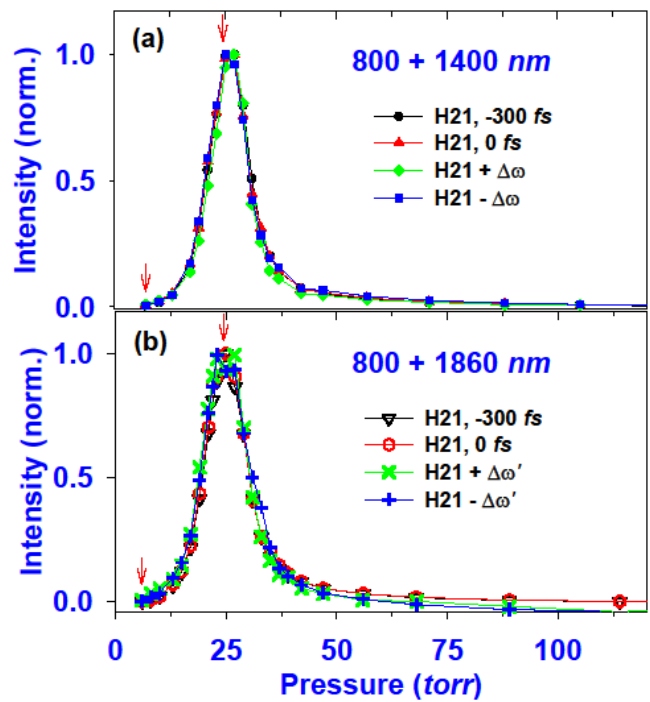

**Figure 3**. Pressure-dependence of the intensities of the harmonic H21 and the mixing fields induced by two-color fields (a) 800 nm and 1400 nm, and (b) 800 nm and 1860 nm. $\Delta\omega = \omega_{800} - \omega_{1400}$, $\Delta\omega' = \omega_{800} - \omega_{1860}$. The red arrows indicate the region where the mixing frequencies are generated with the phase-matched condition.

In the study of the dependence on focus position, the intensities of all fundamental fields are kept constant in the interaction with 25-Torr krypton and the focal point is shifted along the optical axis of the laser fields. Consequently, the intensity of the H21 is built-up with a quadratic effective interaction length (phase-matching condition of HHG [22, 27]) when the focus position is between -1.5 and +3 mm. The negative, zero, and positive values of the focus position here indicate that the focal point is outside, at the exit plane, and inside of the gas cell, respectively. The intensity profiles of the H21 at -300 fs and 0 fs are shown together with that of H21 ± Δω at 0 fs [Fig. 4(a)], and H21 ± Δω' at 0 fs [Fig. 4(b)]. The development of all mixing fields induced either by the driving and the signal pulse or by the driving and the idler pulse are quite similar to that of the main harmonics H21 when the focus position is outside the gas cell ($x \leq 0$) – a signature of small $\Delta k_{mix}$. When $x > 0$, the trends of the H21 ± Δω and the H21 ± Δω' start shifting away from that of the H21. This is clearly revealed in Fig. 4(b) which indicates that the interaction length of the mixing frequencies is longer than that of the main harmonics. The HHG photons are produced at some spatial location along the optical axis within the effective interaction length; these XUV fields consequently participate in the wave-mixing processes in the downstream propagation towards the exit of the same nonlinear medium in the presence of the 800 nm and 1400 nm (800 nm and 1860 nm) laser fields. The different dependence of the intensity of the harmonic and the mixing fields on the interaction length is attributed to cascaded wave-mixing processes where the XUV photons are first generated and then mixed with other photons in the downstream [17, 21]. As the focal point moves deeper into the gas cell, the large phase mismatch makes the intensity of all mixing waves quickly decrease. When $x > +8$ mm, the re-absorption effect of the medium is attributed to the intensity dissipation of all involving fields. The experimental results of these two investigations are thus evidence of the phase-matched generation of the FWM fields under our experimental conditions.

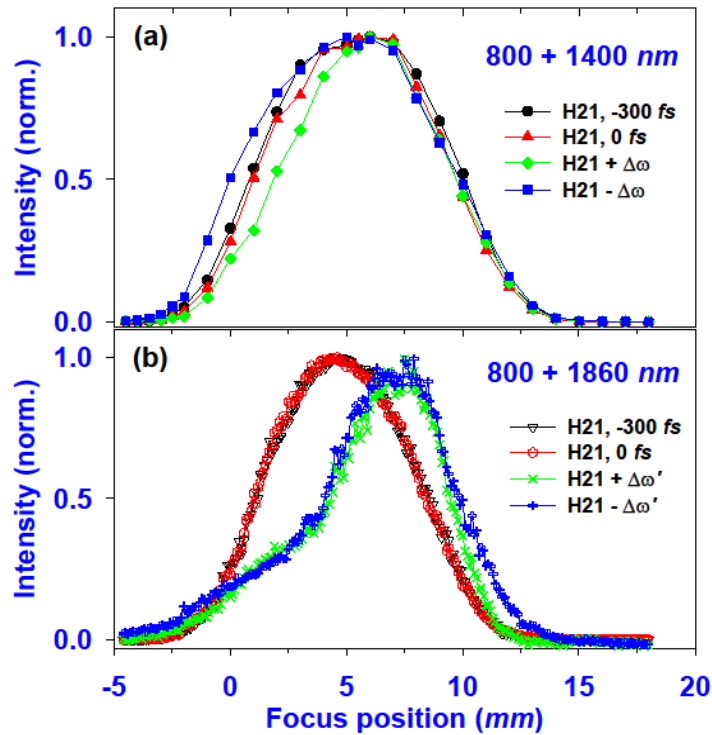

**Figure 4**. Dependence of the intensity of the harmonic H21 and the four-wave mixing fields on focus position for a krypton pressure of 25 Torr. The mixing fields are generated by two-color laser fields (a) 800 nm and 1400 nm, and (b) 800 nm and 1860 nm. $\Delta\omega = \omega_{800} - \omega_{1400}$, $\Delta\omega' = \omega_{800} - \omega_{1860}$.

## 5. Conclusions

We have discussed the perturbative nonlinear optical wave-mixing processes in the XUV region. Four-wave mixing fields induced by mixtures of the two-color laser fields are shown to be produced under the phase-matched condition. The generation of broadband four-wave mixing frequencies with three-color laser pulses has revealed additional evidence of the wave-mixing mechanism in this spectral regime. In the future, a continuum NIR laser can be utilized as a control field to gate an XUV continuum. This might be a promising light source for inner-shell absorption spectroscopy and for the synthesis of bright isolated attosecond pulses when phase-locked control lasers are available. Time evolution of the four-wave mixing fields might also be useful for time-resolved spectroscopy in the XUV range which offers the possibility of studying ultrafast dynamics of core electrons or complex molecules.

## Acknowledgement

Financial support was provided by the ARC Discovery Project Scheme (Grant ID DP170104257).